\documentstyle[aps,twocolumn]{revtex}
\begin{document}
\title{The physical meaning of phase and its importance for quantum teleportation}
\author{S.J. van Enk\\
Bell Labs, Lucent Technologies,
Room 2C-401\\
600-700 Mountain Ave\\
Murray Hill NJ 07974}
\maketitle
\abstract{We argue that the two parties in any quantum teleportation protocol need to share more  resources than just an entangled state and a classical communication channel. 
As the phase between orthogonal states has no physical meaning by itself,
a shared standard defining all relevant phases is necessary. We discuss several physical implementations of qubits and the corresponding physical meaning of phase. }
\section{Introduction}
Quantum teleportation is one of the most important quantum information processing protocols.
It was discovered in 1993 \cite{telep} and several teleportation experiments have been performed since then \cite{martini,dik,akira}. It is well known that the main quantum ingredient is a nonlocal entangled state. It is furthermore assumed that the only additional resource needed is a classical communication channel. Here we will show that that is not quite true. 

In any quantum communication protocol there has to be a well-defined isomorphism between the different Hilbert spaces involved. This is a mathematically trivial requirement, but to establish such an isomorphism in practice may not be trivial. For instance, consider the sender Alice and receiver Bob in a standard teleportation protocol. If Alice is to teleport a qubit in an arbitrary state of the form
\begin{equation}\label{psi}
|\psi_{\theta,\phi}\rangle=\cos(\theta/2)|0\rangle +\sin(\theta/2)\exp(i\phi)|1\rangle
\end{equation}
clearly Bob can only verify this if he and Alice agree on their choice of basis vectors $|0\rangle$,$|1\rangle$ and on the definition of the phase $\phi$. 
How are they going to accomplish this? If they could share a quantum channel during the entire teleportation protocol, Alice could simply send Bob a stream of qubits in, say, the state $|0\rangle$ and a stream of qubits in the state $(|0\rangle+|1\rangle)/\sqrt{2}$. Bob could then measure the qubits and subsequently adjust his basis according to the measurement outcomes. As long as the channel is not so noisy that no error correction can be performed, this would define, in an operational sense, an isomorphism \cite{note}. In fact, a noiseless channel that parallel transports qubits defines what one could call the canonical isomorphism.

In a teleportation protocol, however, we do not allow Alice and Bob to share a quantum channel over and above their entangled state. That is, they of course have to use a quantum channel to establish the entangled state, but they are not allowed to send quantum states later, at the moment the actual quantum teleportation is taking place \cite{note2}. 
Even when Alice and Bob do not share a quantum channel, they still can choose a qubit implementation such that the definition of the basis states is unambiguous. For instance, they can choose $|0\rangle$ and $|1\rangle$ to correspond to zero and one-photon states. Alternatively, they might choose left-hand and right-hand circular polarization states. The latter are defined {\em relative} to the propagation direction of light and as such
can be defined in a consistent way locally in Alice's and Bob's labs. So let's assume they agree on the physical meaning of $|0\rangle$ and $|1\rangle$. The remaining task is then to define the phase $\phi$ appearing in (\ref{psi}). The problem is that the phase $\phi$ has no meaning by itself, because the states $|0\rangle$ and $|1\rangle$ are orthogonal. Thus Alice's and Bob's definitions can be made only relative to some phase standard. For their definitions to be consistent, i.e., to remain the same during the entire teleportation protocol, they are required to {\em share} a phase standard.
Although this shared resource is not a quantum resource, it cannot be replaced by a classical communication channel. For example, one cannot communicate the definition of direction in space over the phone without sharing some additional resource. A classical communication channel transmits just classical bits of information, a phase standard ``transmits'' a physical property.
In the present paper we investigate the nature of such phase standards.

\section{The physical meaning of phase}
We discuss here various proposals for implementing quantum bits (see for example \cite{special}), the meaning of the phase $\phi$ in those cases,
and the phase standards that are needed. 
\begin{enumerate}
\item {\em Spin-1/2 particles.}
One of the most popular representations of a qubit is a spin-1/2 particle, such as an electron or a carbon nucleus.
This is indeed one of the natural implementations of a qubit, as a spin-1/2 system has a two-dimensional Hilbert space associated with its spin. 
The states $|0\rangle$ and $|1\rangle$ are then defined as eigenstates of one component of the angular momentum operator, say, the $z$ component. This of course already assumes Alice and Bob can both define $z$ directions in a consistent way. That is, they may locally define any axis to be the $z$ axis, but their arrangement should be such that the two axes do not rotate with respect to each other. More generally,
since angular momentum generates rotations in space, the definitions of the phase $\phi$ and the states $|0\rangle$ and $|1\rangle$ depend on directions in space. One way to make Alice's and Bob's definitions consistent is to use the fixed stars. This is then an example of a shared resource between Alice and Bob necessary for the ability to teleport reliably.   
\item {\em Photon polarization.}
Another popular representation is photon polarization. Although a photon is a spin-1 particle, it has only two spin (more accurately, helicity) degrees of freedom  because it is massless. Alice and Bob can define their basis states to correspond to left-hand and 
right-hand circular polarization. These two polarization directions can be consistently defined {\em locally}, without a shared standard, because circular polarization states are helicity eigenstates and helicity is defined relative to the propagation direction of the photon. Since the helicity operator generates rotations of the polarization around the propagation direction, however,
the phase $\phi$ depends on the definition of the two spatial directions perpendicular to the propagation direction. Hence, here too, Alice and Bob need to use the fixed stars or a similar resource defining spatial direction. 
\item {\em Photon number.}
One can also encode a qubit in two number states of the electromagnetic field. Given a particular field mode one can choose the states containing 0 and 1
photons to implement $|0\rangle$ and $|1\rangle$. 
In this case, these two states are eigenstates of the (free field) Hamiltonian with different eigenvalues. Since the Hamiltonian generates translations in time, Alice and Bob now need to share a resource that fixes a zero of time. In other words, they need synchronized clocks. Note that even if Alice and Bob have perfect atomic clocks that they synchronized at some point in time, they will still need to keep checking that the clocks don't drift apart due to relativistic or other effects. 
With the time origin fixed, the phase difference between $|0\rangle$ and $|1\rangle$ at that time can be determined locally by measuring any operator that has nonvanishing matrix elements between $|0\rangle$ and $|1\rangle$, such as the electric or magnetic fields.

Note that if one extends the qubit space to higher photon numbers the same arguments apply {\em a fortiori} and consequently the same resource of synchronized clocks is needed. This applies in particular to teleportation with continuous variables. See \cite{rudolph,laser} for related discussions.
\item{\em Harmonic oscillator eigenstates.}
A similar representation makes use of the two lowest vibrational levels of a material particle moving in a one-dimensional harmonic potential \cite{scott}. Again, since these states are eigenstates of the Hamiltonian with different energies, the origin of time must be fixed. The residual phase difference at time zero can be defined locally by measuring position or momentum (both defined relative to the harmonic potential) or any other observable with nonvanishing matrix elements between $|0\rangle$ and $|1\rangle$.

One may also choose {\em degenerate} eigenstates of the 2-D (3-D) degenerate harmonic oscillator so that synchronized clocks are no longer necessary. However, the degeneracy is caused by a symmetry between 2 (3) spatial directions, so that again fixed stars must be used to distinguish the two (three).
\item{\em Atomic energy eigenstates.}
Ground states in atoms or ions  are an experimentally attractive type of implementation. If the atomic or ionic ground state has total angular momentum quantum number $J$, then two states may be chosen out of a multiplet of $2J+1$ degenerate states. Typically, one chooses two eigenstates of the angular momentum along a particular direction, say $J_z$, with eigenvalues differing by one or two units. This is such that one can induce transitions between those two states by using two light beams with either opposite circular polarizations (when the states differ by two units of angular momentum), or with one circular and one appropriate linear polarization (when the states differ by one unit of angular momentum).
This case, therefore, is in an operational sense equivalent to the previous case of photon polarization. 

If one chooses nondegenerate ground states from two different multiplets with different angular momenta, one needs the additional resource of synchronized clocks.
\item{\em Charge and flux states.}
Yet two other types of representations use superconductivity. In particular, one may use charge states (two states with well-defined charge) or magnetic flux states (in fact, states with a well-defined direction of a current) as qubits. These states are not degenerate, and so again synchronized clocks are needed. Since magnetic flux and charge (=electric flux) are complementary variables, the phase at time zero between two charge states can be measured locally by measuring a magnetic flux. Similarly, the phase between flux states can be measured by measuring charge. 
\item{\em General implementations.}
Ignoring practical difficulties one can in principle use any two orthogonal quantum states to implement a qubit. There are, however, some natural restrictions. First, the qubit Hilbert space should be spanned by two eigenstates of the system Hamiltonian, since otherwise the qubit would leave that space at later times.
Preferably these two states should be degenerate, but if synchronized clocks are available then  nondegenerate states can be used as well. If degenerate energy eigenstates are used, then the two basis states can be chosen to be eigenstates of a Hermitian operator $L$ that commutes with the Hamiltonian. In general, the operator commuting with the Hamiltonian corresponds to a symmetry and to ``break'' this symmetry a reference is necessary.
In the examples above the operator $L$ was always an angular momentum operator, exploiting the rotational symmetry of the system. Indeed, this symmetry makes it necessary to share a standard defining direction. 

Another choice, at least in principle, could be to use eigenstates of the momentum operator (assuming the system Hamiltonian is invariant under translations in space) as basic qubit states.  The relative phase between two momentum states may then be defined in terms of absolute position, so that Alice and Bob would need to know (at the very least) their mutual distance. This would not be a practical implementation. Moreover, proper momentum eigenstates are not localized,
so that strictly speaking they cannot even be used locally by Alice and Bob anyway.
\item{\em Relation to Quantum Clock Synchronization}
The previous considerations indicate why quantum clock synchronization based purely on entanglement (see \cite{clock}) is not possible (the argument presented here of course does not invalidate synchronization protocols where a quantum channel is being used to transmit ``ticking'' qubits \cite{ike}). Synchronized clocks are needed to establish an entangled state based on nondegenerate states in the first place. 
Also note that Alice and Bob cannot synchronize their clocks by using an entangled state based on degenerate angular momentum eigenstates (for which they only need to use the fixed stars) and only later perform local operations to lift that degeneracy. The latter operations obviously will introduce only phase differences that depend on the durations of local operations but not on global time differences.
Finally, note that nonlocal entanglement between different types of systems (such as a spin-1/2 particle and a one-dimensional harmonic oscillator) can be obtained only if {\em both} phase standards corresponding to these representations are shared.
\end{enumerate}
\section{Final remarks}
Every quantum communication protocol implicitly assumes that at all times there is a well-defined isomorphism between each of the Hilbert spaces associated with the quantum systems involved in the protocol, even if those systems belong to different parties in different locations. In practice establishing or identifying this isomorphism may not be trivial.
Certain protocols, such as quantum key distribution protocols, involve quantum channels through which the parties are allowed to send unlimited numbers of qubits. In such a case, one party can agree to send the other party a continuous stream of qubits prepared in a predetermined sequence of different pure states. The isomorphism may then be operationally defined and checked during the entire protocol by comparing the outcomes of appropriate measurements performed by the second party against the predetermined sequence. 

For quantum teleportation, however, the use of such a channel is explicitly forbidden, and the isomorphism must be established in an independent way. We argued here that this is only possible if the parties share, 
for the entire duration of the protocol, an appropriate (classical) resource, that goes beyond sharing a classical communication channel. The nature of this resource depends on the physical representation of the qubit. If nondegenerate eigenstates of 
the system's Hamiltonian are used as qubit basis states, then synchronized clocks are necessary.
When degenerate eigenstates are used that are at the same time eigenstates of an operator commuting with the Hamiltonian (corresponding to a symmetry), then a shared classical resource breaking that symmetry is required. Most relevant physical implementations exploit rotational symmetry, and in those cases a resource defining spatial direction, such as the fixed stars, is necessary. 
\section*{Acknowledgments}I thank Chris Fuchs, Norbert L\"utkenhaus, Klaus M\o lmer, Rob Pike, John Smolin and Barbara Terhal for helpful suggestions and comments. 


\begin{thebibliography}{99}
\bibitem{note2}If there is a guarantee that Alice's and Bob's qubit bases do not change in time after they established the isomorphism and the entangled state with their quantum channel, then a phase standard would not be necessary. But of course this begs the question where the guarantee came from.
\end{thebibliography}
\end{document}